# NONLINEAR INTEGRABLE ION TRAPS

S. Nagaitsev

*Fermi National Accelerator Laboratory, Batavia, IL 60510*

V. Danilov

*Spallation Neutron Source Project, Oak Ridge National Laboratory,*

*Oak Ridge, TN 37830*

## Abstract

Quadrupole ion traps can be transformed into nonlinear traps with integrable motion by adding special electrostatic potentials. This can be done with both stationary potentials (electrostatic plus a uniform magnetic field) and with time-dependent electric potentials. These potentials are chosen such that the single particle Hamilton-Jacobi equations of motion are separable in some coordinate systems. The electrostatic potentials have several free adjustable parameters allowing for a quadrupole trap to be transformed into, for example, a double-well or a toroidal-well system. The particle motion remains regular, non-chaotic, integrable in quadratures, and stable for a wide range of parameters. We present two examples of how to realize such a system in case of a time-independent (the Penning trap) as well as a time-dependent (the Paul trap) configuration.

In the oblate ($\xi, \eta, \phi$) and prolate ($u, v, \phi$) spheroidal coordinates, any axially symmetric potential, $V(\xi,\eta)$ or $U(u,v)$, that results in separability of the Hamilton-Jacobi equation must have the following form:

$$V(\xi,\eta) = \frac{f_1(\xi) + f_2(\eta)}{\xi^2 + \eta^2}; \quad U(u,v) = \frac{g_1(u) + g_2(v)}{u^2 - v^2}, \qquad (1)$$

where $f_1$, $f_2$, $g_1$ and $g_2$ are arbitrary functions. The potentials in the form (1) satisfying the Laplace equation (and having no singularity at the origin) are the so-called Vinti potential [1] in oblate and the Two-Fixed-Coulomb-Centers (2FCC) potential [2] in prolate coordinates respectively. These two potentials can be presented as follows:

$$V(x,y,z) = \frac{\sigma_1}{a} \frac{\left(4a^2 - \left[\sqrt{(a+r)^2 + z^2} - \sqrt{(a-r)^2 + z^2}\right]^2\right)^{1/2}}{\sqrt{(a+r)^2 + z^2}\sqrt{(a-r)^2 + z^2}}, \qquad (2)$$

$$U(x,y,z) = \frac{\sigma_2}{\sqrt{(z-c)^2 + r^2}} + \frac{\sigma_3}{\sqrt{(z+c)^2 + r^2}}, \qquad (3)$$

where $\sigma_1$, $\sigma_2$, $\sigma_3$, $a$ and $c$ are arbitrary parameters, and $r^2 = x^2 + y^2$.

Let us now consider the particle motion in an ideal Penning trap [3]. The Hamiltonian for a particle in cylindrical coordinates is given by

$$H = \frac{1}{2m}\left(p_z^2 + p_r^2\right) + \frac{p_\phi^2}{2mr^2} - \frac{p_\phi \omega_c}{2m} + \frac{m\omega_c^2 r^2}{8} + \frac{eV_0}{4d^2}\left(2z^2 - r^2\right), \tag{4}$$

where $e$ is the ion charge, $\omega_c = \frac{eB}{mc}$ is its cyclotron frequency, $m$ is its rest mass, $c$ is the speed of light, $d$ is the trap dimension and $V_0$ describes the axially symmetric quadrupole electrostatic potential, $\varphi = \frac{V_0}{4d^2}\left(2z^2 - r^2\right)$, of a Penning trap. First, it is obvious that the angular momentum, $p_\phi$, is the integral of motion and, thus, a conserved quantity. Second, the trap voltage, $V_0$, can be chosen such that

$$V_0 = \frac{mc^2}{e} \frac{\omega_c^2 d^2}{6c^2}. \tag{5}$$

For example, for a trapped proton in a $B = 1\,\text{T}$ magnetic field and $d = 1\,\text{cm}$, the Eq. (5) yields $V_0 \approx 1.6\,\text{kV}$. With such a voltage the effective trap potential becomes

$$U_{eff} = \frac{p_\phi^2}{2emr^2} - \frac{p_\phi \omega_c}{2em} + \frac{V_0}{2d^2}\left(z^2 + r^2\right). \tag{6}$$

One can easily verify that the potential (6) satisfies both forms of Eq. (1) and it can be, therefore, combined with either the Vinti (2) or the 2FCC (3) potentials. The Hamilton-Jacobi equation in such a combined potential is separable in both the prolate and oblate spheroidal coordinates and, thus, the motion is integrable in quadratures. Figures 1 and 2 present the contour plot of the effective potential Eq. (6) combined with the Vinti and the 2FCC potentials for $p_\phi = 0$, $a = 1$ and $c = 1$.

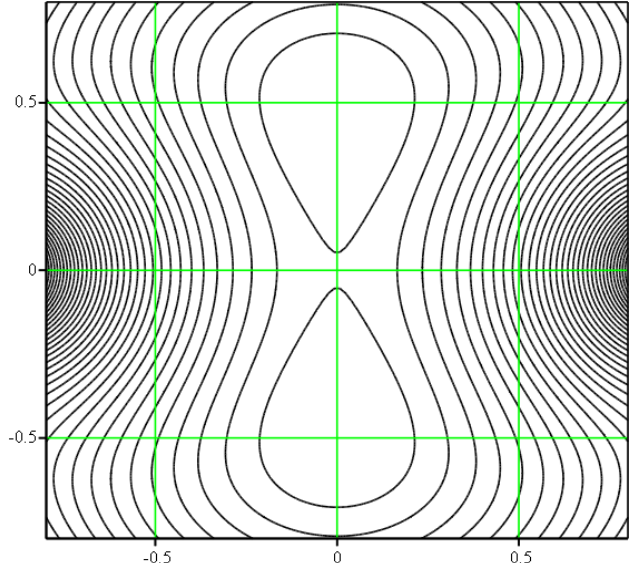

**Figure 1:** The contour plot of the effective potential Eq. (6) combined with the Vinti potential for $p_\phi = 0$ and $a = 1$ in the *x-z* plane. One can see the singularity at $r = a$. The value of the parameter $\sigma_1$ is chosen such that the potential has a double-well shape along the vertical (*z*) axis.

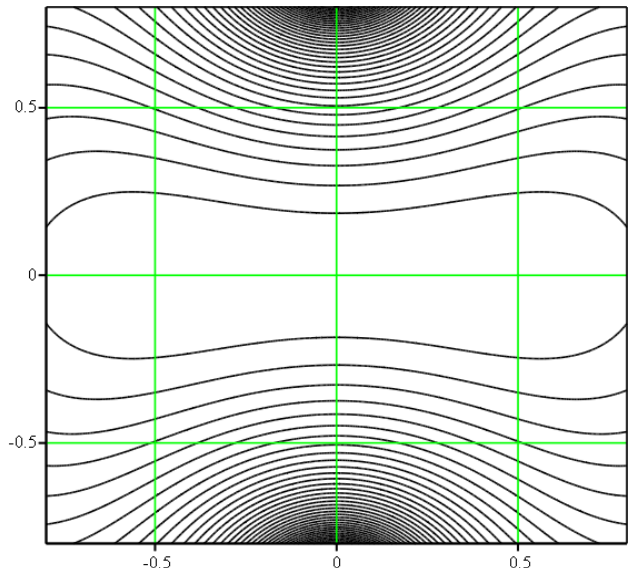

**Figure 2:** The contour plot of the effective potential Eq. (6) combined with the 2FCC potential for $p_\phi = 0$ and $c = 1$ in the *x-z* plane. One can see the singularities (the Coulomb Centers) at $z = \pm c$. The values of the parameters $\sigma_2$ and $\sigma_3$ are chosen to be equal and such that the potential has a toroidal-well shape along the horizontal (*x*) axis.

We will now turn our attention to the Paul trap [4]. In an ideal axially symmetric Paul trap the particle Hamiltonian is time-dependent

$$H = \frac{1}{2m}\left(p_x^2 + p_y^2 + p_z^2\right) + \frac{eW(t)}{4d^2}\left(2z^2 - x^2 - y^2\right), \tag{7}$$

where $W(t)$ is the time-dependent trap voltage amplitude. Normally, this voltage is a sine-wave function, $W_0 + W_1 \sin(\omega t)$. However, in this paper we shall only assume that this voltage is a periodic function of time, $W(t) = W(t+T)$, and can be controlled to attain certain properties described below. The equations of motion are:

$$\begin{cases} \ddot{x} + K_x(t)x = 0 \\ \ddot{y} + K_y(t)y = 0, \\ \ddot{z} + K_z(t)z = 0 \end{cases} \tag{8}$$

where $K_x = K_y = -\frac{eW(t)}{2md^2}$ and $K_z = \frac{eW(t)}{md^2}$ are periodic focusing functions. Equations (8) are the uncoupled Hill's equations. Such an equation was first solved by Ermakov [5], who obtained its integral of motion. The motion, described by (8), is linear and periodic. Thus, the Paul trap's Hamiltonian has three Ermakov integrals of motion. In this letter we are interested in finding an additional time-dependent potential, $Q(x, y, z, t)$, which retains the Hamiltonian's integrabilty but makes the particle motion non-linear. Reference [6] describes a method, developed for a 2d particle motion in accelerators, to obtain such integrable non-linear potentials. In this Letter we will closely follow it and apply it to a 3d time-dependent trap. First, let us assume that $K_x = K_y = K_z = K(t) = K(t+T)$. Clearly, one cannot attain such focusing functions continuously with a quadrupole potential, however we will demonstrate below how to obtain them periodically, during part of the time, by choosing an appropriate voltage function, $W(t)$. The Hamiltonian has the form,

$$H = \frac{1}{2m}\left(p_x^2 + p_y^2 + p_z^2\right) + \frac{mK(t)}{2}\left(x^2 + y^2 + z^2\right) + Q(x, y, z, t), \tag{9}$$

where the unknown potential $Q$ satisfies the stationary Laplace equation, $\Delta Q = 0$. Following the Ermakov's method [5], we introduce an auxiliary function, $\beta(t)$, which is the solution to the following differential equation:

$$\frac{d^2\left(\sqrt{\beta}\right)}{dt^2} + K(t)\sqrt{\beta} = \frac{1}{\sqrt{\beta^3}}. \tag{10}$$

Then a new unitless "time" variable, $\psi$, can be introduced, such that

$$\frac{d\psi}{dt} = \frac{1}{\beta(t)}. \tag{11}$$

One can then make a canonical variable transformation to the so-called normalized variables. The new Hamiltonian in these new variables is

$$H_N = \frac{p_{xN}^2 + p_{yN}^2 + p_{zN}^2}{2} + \frac{x_N^2 + y_N^2 + z_N^2}{2} + \beta(\psi) Q\left(x_N\sqrt{\beta(\psi)}, y_N\sqrt{\beta(\psi)}, z_N\sqrt{\beta(\psi)}, t(\psi)\right), \tag{12}$$

where

$$q_N = \frac{q}{\sqrt{\beta(t)}},$$
$$p_N = \frac{p}{m}\sqrt{\beta(t)} - \frac{\dot\beta(t) q}{2\sqrt{\beta(t)}}, \tag{13}$$

and $q$ stands for either $x$, $y$ or $z$ and $p$ is similarly either $p_x$, $p_y$ or $p_z$. Below are the two main ideas of this method:

1. If the potential $R(x_N, y_N, z_N) = \beta(\psi) Q(x_N\sqrt{\beta(\psi)}, y_N\sqrt{\beta(\psi)}, z_N\sqrt{\beta(\psi)}, t(\psi))$ in Eq. (12) is independent of the time variable, $\psi$, then the system has at least one integral of motion, namely the Hamiltonian $H_N$ itself in the new variables (13). Note that the potential $R$ satisfies the stationary Laplace equation, if $Q$ does.

2. The potential of a spherically symmetric oscillator, $x_N^2 + y_N^2 + z_N^2$, is separable in ellipsoidal (as well as in several other) coordinates. If, in addition, the potential $R$ belongs to a class of potentials (described in Ref. [7]) integrable in ellipsoidal coordinates, then the Hamiltonian (12) is also separable in ellipsoidal coordinates. Such potentials are generally non-linear and have several adjustable parameters. A sub-class of these non-linear potentials is the described above axially-symmetric potentials, the Vinti (2) and the 2FCC (3).

Thus, we now have two remaining tasks: to demonstrate (1) how to make the potential $R$ time-independent and (2) how to provide equal focusing functions, $K(t)$. We will use the 2FCC potential as an example. For the potential $R$ to be time-independent it is easy to show that the potential $Q$ must be of the form,

$$Q(x,y,z,t) = \frac{1}{\sqrt{\beta(t)}} \left( \frac{\sigma_2}{\sqrt{\left(z - c_1\sqrt{\beta(t)}\right)^2 + r^2}} + \frac{\sigma_3}{\sqrt{\left(z + c_1\sqrt{\beta(t)}\right)^2 + r^2}} \right). \tag{14}$$

To demonstrate how to provide equal focusing functions we will follow the Ref. [6]. Let us divide the period $T$ into two portions $T_0$ and $T_1$. During $T_0$, only the quadrupole potential, $W(t)$, can be nonzero. During $T_1$, the quadrupole potential is off and only the nonlinear potential, $Q$, Eq. (14) is nonzero. The key to this idea is to make the function $\beta(t)$, associated with each of the equations (8), equal for $x$, $y$ and $z$ coordinates only during the time period $T_1$, whereas during the period $T_0$ the functions can be different. The additional special conditions for the quadrupole focusing functions during the period $T_0$ can be written as follows: at the beginning (subscript $b$) and at the end (subscript $e$) of $T_0$, $q_e = q_b$ and $p_e = p_b - kq_b$. This is a well-known linear focusing transformation corresponding to a thin spherically-symmetric lens. The fact that such a transformation exists with focusing functions in Eq. (8) can be easily proven. In practice, to realize such a complicated time dependence of various potentials one can use multiple electrodes, each powered by its own power supply. For an axially-symmetric case such electrodes could be rings, similar to a planar surface-electrode Paul trap.

In conclusion, we have presented two types of nonlinear integrable ion traps. The nonlinear potentials can be introduced both as stationary potentials (electrostatic plus a uniform magnetic field) and as time-dependent electric potentials. The particle motion in such traps is (by definition of integrability) regular and non-chaotic.

This research is supported by UT-Battelle, LLC and by FRA, LLC for the U. S. Department of Energy under contracts No. DE-AC05-00OR22725 and DE-AC02-07CH11359 respectively.